\documentclass[aps,prl,onecolumn,showpacs]{revtex4}
\usepackage{amssymb}
\usepackage{xcolor}
\usepackage{graphicx}
\usepackage{amsmath}

\begin{document}

\title{ Triplet odd-frequency  superconductivity in hybrid superconductor-ferromagnet structures}

\author{F. Sebastian~Bergeret}
\affiliation{
Centro de F\'{i}sica de Materiales (CFM-MPC), Centro Mixto CSIC-UPV/EHU, Manuel de Lardizabal 5, E-20018 San Sebasti\'{a}n, Spain}
\affiliation{Donostia International Physics Center (DIPC), Manuel de Lardizabal 4, E-20018 San Sebasti\'{a}n, Spain}

\author{Anatoly F. Volkov}
\affiliation{Theoretische Physik III, Ruhr-Universit\"at Bochum, D-44780 Bochum, Germany}

\begin{abstract}
We present an overview of the contributions made by  Konstantin B. Efetov and the authors to the field of triplet odd-frequency superconductivity in hybrid ferromagnet-superconductor structures\cite{bergeret2005odd} and the repercussions they had on experimental and theoretical research.  Konstantin B. Efetov passed away in August 2021. We hope the present short review gives a faithful testimony of an important part of his scientific legacy.
\end{abstract}

\maketitle
\affiliation{Theoretische Physik III,\\
Ruhr-Universit\"{a}t Bochum, D-44780 Bochum, Germany}
\date{\today }

\section{Introduction}

Shortly after the appearance of the fundamental BCS theory of
superconductivity \cite{BCS57}, several papers were published in which
the action of an exchange or Zeeman field on the spins of Cooper
pairs were studied.  Abrikosov and Gor'kov considered the influence of
paramagnetic, randomly distributed impurities on superconductivity and
predicted,  on the one hand, a strong suppression of superconductivity,  and on the other hand, a gapless state  in
certain range of magnetic impurities  concentration\cite{abrikosov1960contribution}. 
Important
results were obtained in the seminal works  by Fulde,Ferrel and Larkin,
Ovchinnikov \cite{fulde1964superconductivity,larkin1965nonuniform} who analyzed the superconducting state
in the presence of a homogeneous spin-exchange field.
They predicted a transition from the uniform superconducting state into a
nonhomogenous state, the so-called LOFF-state.  Such inhomogeneous state has being intensively
investigated in different contexts, as heavy fermions superconductivity\cite{matsuda2007fulde}, ultracold gases \cite{kinnunen2018fulde}, and quantum chromo-dynamics\cite{casalbuoni2004inhomogeneous}.
In conventional superconductors, non-magnetic impurities narrow the range of the FFLO state\cite{aslamazov1969influence} in the phase diagram, making its detection extremely difficult. 

However, a state with an inhomogeneous Cooper pair density state may appear 
in a  different platform which is superconductor-ferromagnet (S/F)  layered structures.  
The superconducting correlations generated via the proximity effect 
interact with the effective exchange field  $h$ of the ferromagnet.
The action of the exchange of $h$ leads to spatial oscillations of the superconducting correlations (Cooper pairs)  in F, which in turn leads to striking effects. One of them is the realization of the so-called $\pi $ - state
in an S/F/S Josephson junction, in which the ground state corresponds to a $\pi$ 
phase difference between the superconductors\footnote{ {Note that the possibility of a  $\pi$-junction was first predicted first predicted in Refs. \cite{kulik1966magnitude,bulaevskii1977superconducting} 
in which an SIS tunnel junction with magnetic impurities in an insulating layer I was consider. In this article we focus on metallic SFS Josephson junction in which the long-range triplet superconducting correlations are induced (see section below). }}. This manifests as  
a change of the sign of the Josephson critical current in a certain interval of temperatures and exchange energies as predicted in Refs. \cite{buzdin1982critical,buzdin1991josephson} and verified experimentally in \cite{ryazanov2001coupling,kontos2002josephson,sellier2003temperature,weides2006high}. The spatial oscillation of the superconducting correlations also manifests as a non-monotonic behavior of $T_c$ \cite{buzdin1990transition,jiang1995oscillatory,lazar2000superconductor} and the density of states\cite{buzdin2000density,kontos2001inhomogeneous} of  S/F bilayers as a function of the F-layer thickness.

It is important to note that until 2001, all work on S/F structures assumed homogeneous magnetization. Under these circumstances, the Copper pairs  {decay exponentially in the  ferromagnetic layer over  a length of the order  {$\xi _{h}\approx \sqrt{D_F/h}$, where $D_F$} is the diffusive coefficient of F, even though mesoscopic fluctuations can increase this length, turning the exponential decay into a slower one \cite{zhou1998superconducting,zyuzin2000theory}}.  Typical ferromagnets have exchange fields of the order of eV. Hence, $\xi _{h}$ is extremely small $\lesssim 1$nm. This is why the transition to the $\pi$-state could only be observed in experiments on SFS junctions with ferromagnetic alloys ({\it e.g. }, Cu$_x$Ni$_{1-x}$ ) containing very low concentrations of ferromagnetic atoms. A new twist appeared with the work of Konstantin B. Efetov and the authors. They showed that if the magnetization of the ferromagnet is inhomogeneous, then Cooper pairs in the triplet state and with spin parallel to the local magnetization of F  can be created\cite{bergeret2001long,bergeret2005odd}. These pairs can penetrate the ferromagnet over long distances of the order of  {$\sqrt{D_F/T}$}. Such a long-range proximity effect  has been observed in numerous experiments \cite{keizer2006spin,robinson2010controlled,khaire2010observation,anwar2010long,sosnin2006superconducting,anwar2011inducing,anwar2012long,salikhov2009experimental,blamire2014interface,banerjee2014evidence,di2015signature,martinez2016amplitude,massarotti2018electrodynamics,niedzielski2018spin,caruso2019tuning,aguilar2020spin,ahmad2020electrodynamics,fermin2022superconducting} and opened an active research field\cite{eschrig2011spin,linder2015superconducting,eschrig2015spin}.

In this article, we provide a brief review of the triplet odd-frequency superconductivity theory, its impact on the research community, and more recent developments.

\section{Homogeneous magnetization}

As mentioned in the introduction, the first theoretical works on S/F structure assumed an homogeneous magnetization of the F layer. First Buzdin and Kupriyanov Ref.\cite{buzdin1991josephson} analyzed a S/F/S
Josephson junction in the dirty limit on the basis of the Usadel equation. At that time equations were not written in the spin space. With today's perspective, we know that it is convenient to keep the  structure in spin space.in order to understand the fundamentals of  triplet odd-frequency superconducting correlations. 

If we assume a weak proximity effect the Usadel  equation is a diffussion equation for  the  quasiclassical anomalous  Green's function $\hat{f}(x)$ in the
F film, describing the diffusion of pairs:
\begin{equation}
-\partial _{xx}^{2}\hat{f}+\kappa _{\omega }^{2}\hat{f}+i  {\rm sgn \omega_n \frac{\kappa _{h}^{2}}{2}}\left\{
\sigma_3,\hat{f}\right\}=0\text{,}  \label{1}
\end{equation}%
where $\kappa _{\omega }^{2}=2|\omega_n |/D_{F}$,  {$\omega_n =\pi T(2n+1)$ }is the
Matsubara frequency, $D_{F}$ is the diffusion coefficient in F,  {$\kappa
_{h}^{2}=(h/D_{F})$}. The matrix $\sigma_3$ is the third Pauli matrix in spin space. { For the the matrix describing the anomalous correlations $f_{ss´}\sim\langle c_s c_{s'}\rangle$   we use the representation proposed  in Ref. \cite{ivanov2006minigap}.
In Eq. (\ref{1}), } we assume that the exchange field is parallel to the magnetization which is homogeneous pointed in $z$ direction.
The last term Eq.(\ref{1}) describes such exchange field, the expression in the curly brackets is an anticommutator. 

Eq.(\ref{1}) is supplemented by the boundary conditions \cite{kuprianov1988influence}
\begin{equation}
-\partial _{x}\hat{f}|_{x=0}=\kappa _{b}F_{S}\text{,}
\label{2}
\end{equation}%
where the coefficient $%
\kappa _{b}=(R_{b\Box }\sigma _{F})^{-1}$ characterizes the
transparency of the S/F interface, $R_{b\Box }$ and $\sigma _{F}$ are the
S/F interface resistance per unit area and the conductivity of the
ferromagnet. The function at the right-hand side of Eq.(\ref{2}) is the condensate function in conventional S layer. It is in the singlet state and hence, in the present representation, proportional to the unit matrix $\sigma_0$. 
It  has the standard BCS form
\begin{equation}
F_{S}=\frac{\Delta }{\sqrt{\omega _{n}^{2}+\Delta ^{2}}}\text{,}  \label{3}
\end{equation}

Already at that simple level one can draw interesting conclusions from Eqs.(\ref{1}-\ref{2}). 
First, in the absence of an exchange field, the condensate function $\hat f$ is an scalar, {\it i.e.} is proportional to the unit matrix in the spin-space. This scalar  corresponds to the singlet component. Hence, as expected,  the proximity effect in a metal without exchange field can only induce singlet pair correlations. 
In contrast, in the presence of an exchange field, one can easily check that the general form of $\hat f$ is :
\begin{equation}
    \hat f = f_s {\hat \sigma_0}+f_3 {\hat \sigma_3};,
    \label{eq:f_hom}
\end{equation}
where $f_s$ describe the singlet and $f_3$ the triplet component of the condensate. 
The anticommutator in the third term of Eq. (\ref{1}) couple these two components. In other words, the presence of an exchange field leads always to a singlet-triplet conversion. Since all this is in the diffusive limit, where the function $\hat f$ does not depend on momentum, the condensate has an  s-wave symmetry, and therefore the triplet component has to be odd in Matsubara frequency to satisfy the Pauli exclusion principle \cite{bergeret2001long}. This can be checked directly from Eq. (\ref{1}).

The odd-frequency triplet Cooper pairs always exist in uniform superconductors with a build-in exchange field \cite{matsuda2007fulde,kinnunen2018fulde,casalbuoni2004inhomogeneous}. However in this case both components, singlet and triplet, coexist so that the contribution of the triplet component to observable quantities is masked by the singlet component.  There are  proposals of having triplet odd-frequency superconductivity  in solids (see a review \cite{linder2019odd} and references therein).  These proposals are based on the ideas of Berezinskii in the context of $^3$He\cite{berezinskii1974new}, who considered a system with retarded interaction and showed that an odd-frequency triplet superfluidity is possible in such a system.  However, triplet odd-frequency in all these proposal remained only a hypothetical state.


The solution  of Eqs.(\ref{1}-\ref{2}) can be written straightforwardly: 
\begin{equation}
\hat{f}(x)=2\text{Re}A_{0}\exp (-\kappa x) {\hat \sigma_0}+2i\text{%
Im}A_{0}\exp (-\kappa x) {\hat \sigma_3} \text{,}  \label{4}
\end{equation}%
where $A_{0}=(\kappa _{b}/\kappa )F_{S}$
and $\kappa ^{2}=\kappa _{\omega }^{2}+   {i {\rm sgn}\omega_n \kappa _{h}^{2}}$. Thus, both the singlet and triplet components decay and oscillate 
over the same length scale. Because the magnetization of the ferromagnet is homogenoues only the triplet
component with zero projection of the total spin is generated. If the exchange field is large enough, as in usual ferromagnets, the triplet component, as the singlet, decays over a short length
of the order of $\xi _{h}\approx \sqrt{D_{F}/h}$.  The  oscillatory behavior of the condensate  leads to oscillations of the critical
Josephson current $I_{c}(h,T)$ in S/F/S junction \cite{buzdin1982critical,buzdin1991josephson} described in the introduction. 
\begin{figure}
    \centering
    \includegraphics[width=1 \textwidth]{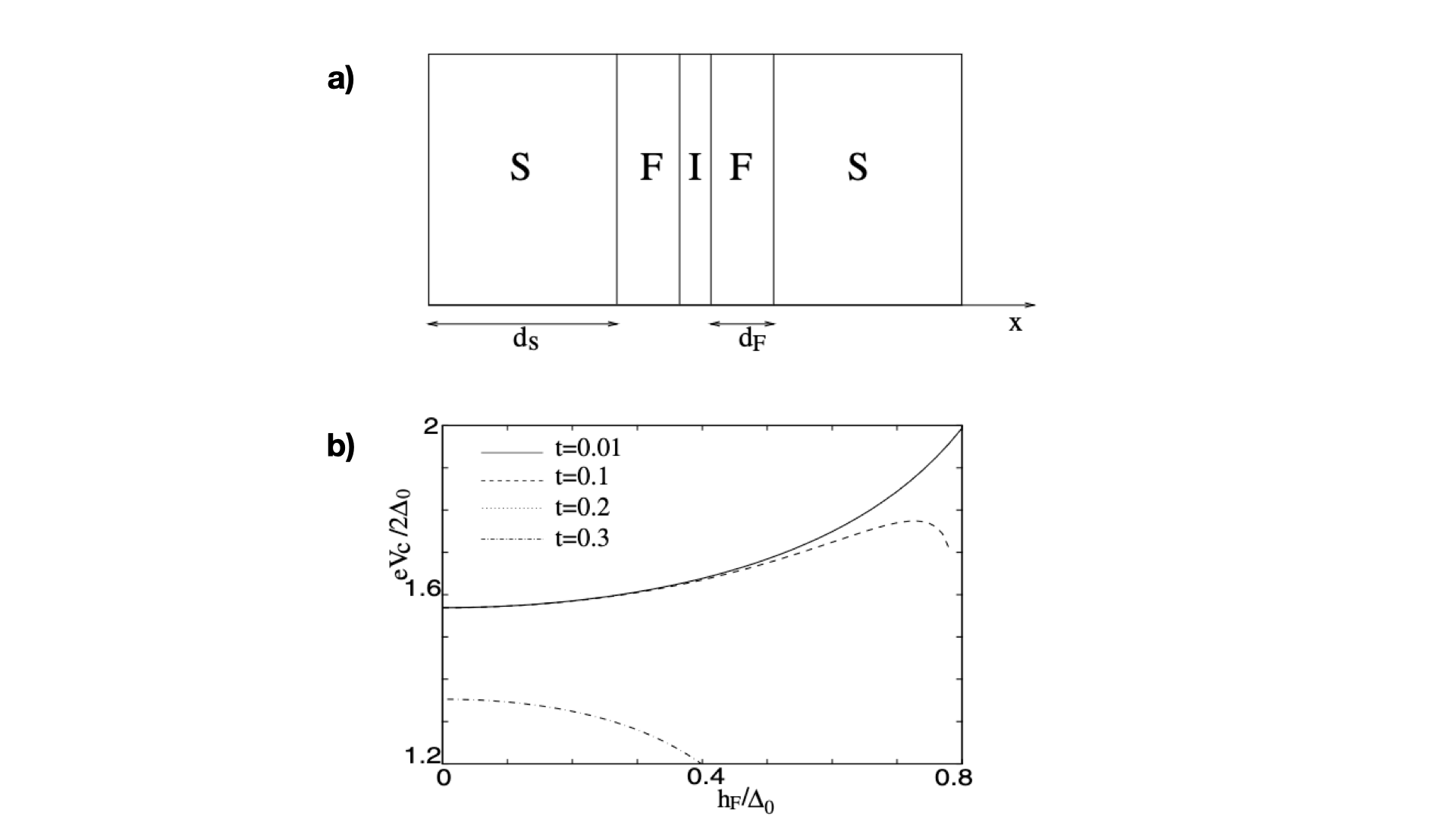}
    \caption{ a) The Jospehson junction analyzed in our work \cite{bergeret2001enhancement}. b) The normalized critical current as a function of the exchange field in teh antiparllel configuration of junction a).  
    Adapted from Ref. \cite{bergeret2001enhancement}. Copyright (2001) by the American Physical Society.}
    \label{fig1}
\end{figure}

An interesting result,  in the  case of co-linear magnetization, was obtain by our group in 2001\cite{bergeret2001enhancement}.  The inclusion of ferromagnetic layers F into a tunnel\
S/I/S ($I$
is  a thin insulating layer) Josephson junction may lead to an increase of the  critical  current $I_{c}$. Specifically,  we studied the tunnel SF/I/FS junction sketched in Fig. \ref{fig1}a, 
and showed that the critical current  increases in case of
antiparallel magnetisations between   the left and right  F films F.  The critical current $I_c^{(ap)}$ in this case 
may
 exceed the current $I_{c}$ in a  S/I/S junction without F layers.  
Explicitly,  we found  that 
\begin{equation}
\frac{I_{c}^{ {(ap)}}(h)}{I_{c}(0)}=2\pi T\sum_{\omega _{n}}\frac{%
\Delta (T)}{\sqrt{(\omega _{n}^{2}+\Delta ^{2}(T,h)-h^{2})^{2}+4\omega
_{n}^{2}h^{2}}}  \label{5}
\end{equation}%
where $\omega _{n}=\pi T(2n+1)$. 
 Figure \ref{fig1}b shows  the dependence of normalized critical current as a function of the exchange field in  the antiparallel configuration. A plausible experimental confirmation of that prediction was presented  in Ref.\cite{robinson2010enhanced}. 

Another effect in S/F structures with an uniform magnetization   predicted  by us is  induced
magnetic moment in the superconductor S via the inverse proximity effect\cite{bergeret2004spin,bergeret2004induced}
 {This effect is a direct consequence of the triplet component induced in the F layer which via the inverse proximity effects can  penetrate back  into the S region. It is an equilibrium effect,  a consequence of the proximity effect, and therefore  very different  from the spin injection in superconductors under non-equilibrium conditions\cite{aronov1976spin,bergeret2018colloquium}}.

\begin{figure}
    \centering
    \includegraphics[width=.65 \textwidth]{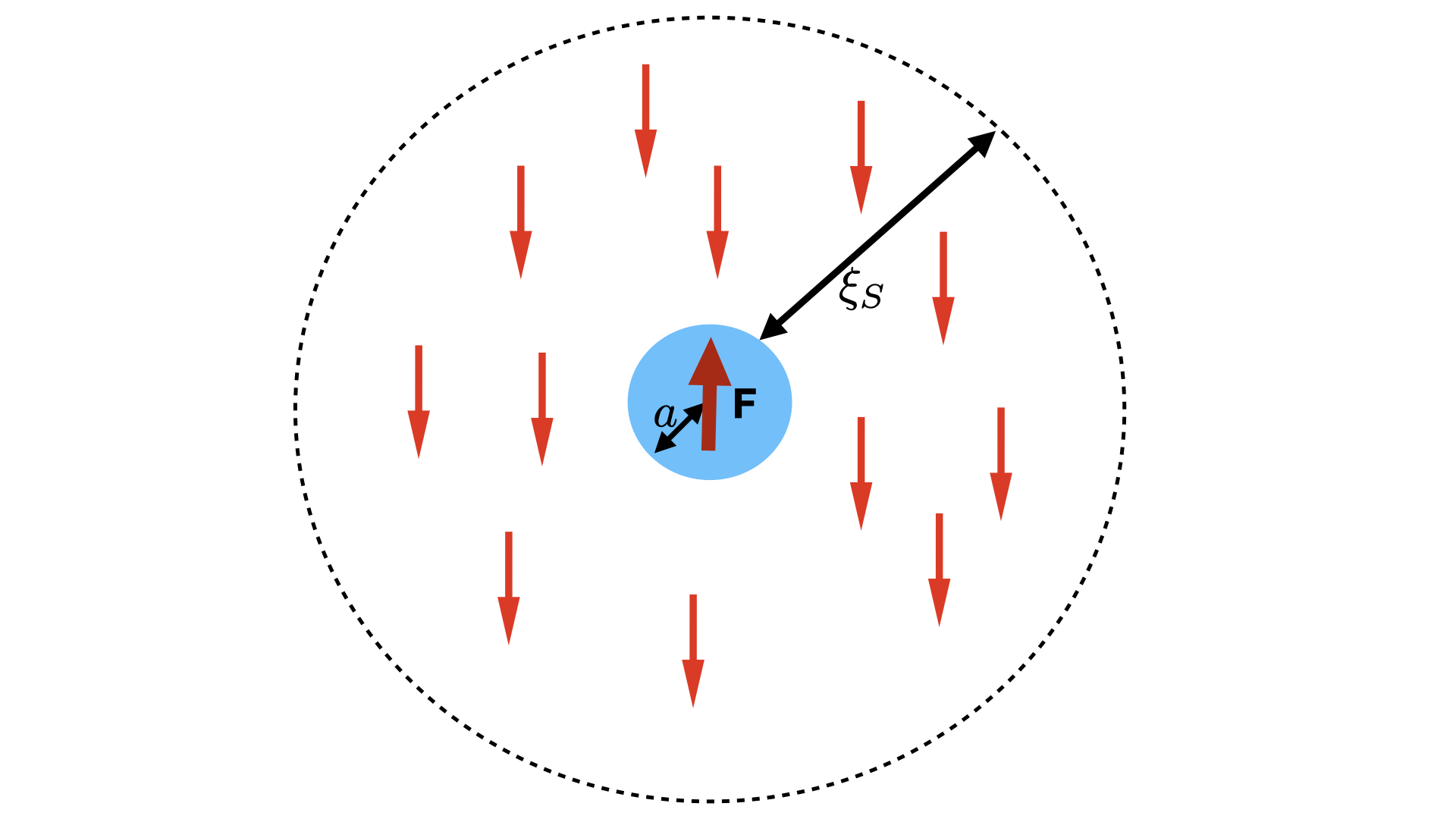}
    \caption{ Schematic representation of the screening effect of a magnetic particle embedded in a superconductor, via the inverse proximity effect.  }
    \label{fig2}
\end{figure}

In  diffusive systems 
 the induced magnetic moment,  $\mathbf{M}_{S}$,  has a direction opposite to the magnetization in $%
\mathbf{M}_{F}$, leading to the   screening of the spin of a magnetic particle embedded in a superconductor,as  sketch in  \ref{fig2}. The magnetic moment in $S$ extends over the superconducting coherent length. 
Experimental evidences  of
the magnetism induced in  S  by an adjacent ferromagnetic F has been presented  in
Refs.\cite{xia2009inverse,salikhov2009experimental,satariano2021inverse}

\begin{figure}
    \centering
    \includegraphics[width=1 \textwidth]{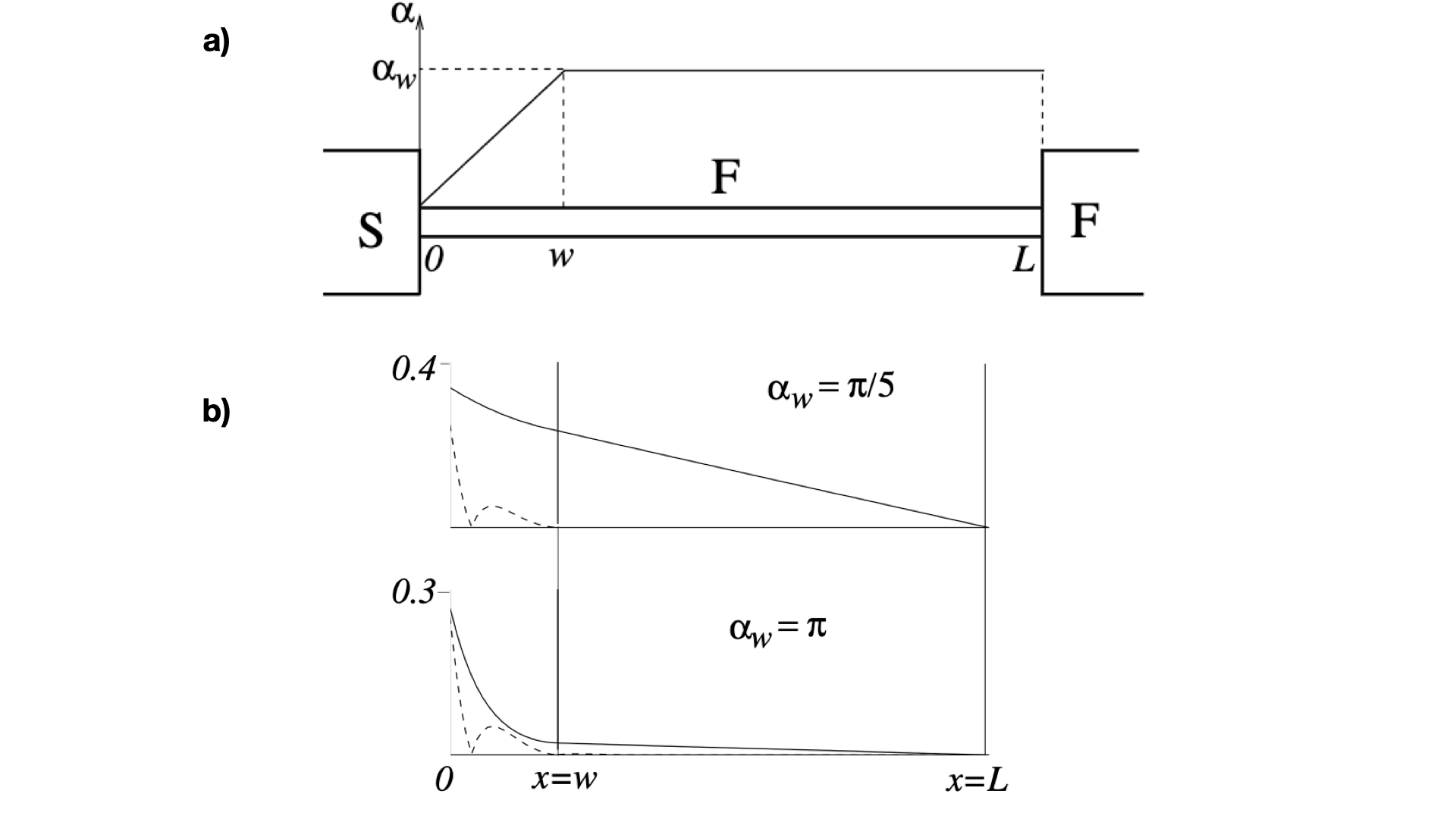}
    \caption{ a) Schematic view of the structure consider in our first work predicting the long-range triplet component\cite{bergeret2001long}. A superconducting electrode is attached to a ferromagnet. In our original work a Bloch domain wall occupies the region $0<x<w$. Whereas an homogenoeus magnetization is assumed at $x>w$.  In the main text we consider a simpler situation in which the magnetization of the region $0<x<w$ is homogeneous but perpendicular to the one in $x>w$. b) The calculated singlet (dashed lines) and long-range triplet (solid line) components assuming a Bloch domain wall described by a vector ${\bf n}=(0,\sin Qx,\cos Qx)$.  Here $\alpha_w=Qw$. 
    Adapted  from Ref.  \cite{bergeret2001long}. Copyright (2001) by the American Physical Society.}
    \label{fig3}
\end{figure}

\section{Inhomogeneous magnetization: Long-range triplet component}
The physics described in the previous section turned out to be much richer when the  magnetization  of the ferromagnet is spatially inhomohgenoeus. This is case of ferromagnets with domain wall, or multiyared magnetic structures. 
Our  2001  work Ref. \cite{bergeret2001long} 
predicted an induced  long-range triplet component  in ferromagnet with a domain wall. Such component may penetrate the F region   
on a distance, $\xi _{T}\approx $ $\sqrt{D_{F}/T}$, which can be much longer than  the length $\xi _{h}$ \footnote{Shortly after our work \cite{bergeret2001long}, the authors of Ref. \cite{kadigrobov2001quantum} considered the possibility of long-range triplet component  induced in a quasiballistic  domain wall.  This case is more complicated and requires solving the Eilenberger equation. The first  detailed study of this case in the context of long-range triplet component was carried out in Ref.  \cite{volkov2008odd}}.  
Specifically,  we consider a S/F structure as the one sketched in Fig.\ref{fig3}a with a Bloch domain wall with in the region$0<x<w$.
By solving the Usadel equation we found that beside the discussed  singlet and short-range triplet components, the inhomogenous magnetization of  the domain wall  induces a long-range triplet component which may penetrate into the  homogeneous domain  over the length $\xi_T$. This finding provided a plausible explanation for   experimental observations of a long-range proximity effects in S/F mesoscopic structures which so far  could not be explained by the existing theories \cite{petrashov1999giant,giroud1998superconducting,aumentado2001mesoscopic}.  Moreover, a natural consequence of our theory 
is the prediction of a long-range Josephson effect in SFS junctions with inhomogeneous magnetization\cite{bergeret2001josephson}. 

To describe systems with inhomogenoeus magnetization one needs to generalize  Eq. (\ref{1}) for an  exchange field with arbitrary  direction, ${\bf h}=h{\bf n}$,  where ${\bf n}$ is a unit vector pointing on the direction of the local exchange field: 
\begin{equation}
-\nabla ^2 \hat f + \frac{1}{2}\{ \kappa_\omega^2+i  {{\rm sgn}\omega_n}\kappa_h^2 \mathbf{n \cdot {\boldmath \sigma}} ,\hat f\}=0\; .
\label{eq:linUsadelcompl}
\end{equation}
In the absence of a superconducting phase,  the condensate function $\hat f$ has only a spin structure, similar to Eq. (\ref{eq:f_hom} but with all triplet components:
\begin{equation}
    \hat f = f_s {\hat \sigma_0}+{\bf f_t}{\boldmath{\cdot  {\hat{\boldsymbol{\sigma}}}}};. \label{eq:gen_f}
\end{equation}
where  ${\bf f_t}$ is the triplet vector with odd-frequency components. 
To illustrate the appearance of the long range triplet component we consider a simple case: 
We assume that in the region $0<x<w$, Fig. \ref{fig3}a, the  exchange field points in $x$ direction, such that  $\mathbf{n}=(1,0,0)$, whereas for $x>w$ the exchange points in $z$ direction. In this case the triplet vector has the form  ${\bf f_t}=(f_1,0,f_3)$. 
The $f_1$ component is generated in the $0<x<w$ region by the singlet-triplet conversion 
described by the last term in Eq. (\ref{eq:linUsadelcompl}), whereas $f_3$ is generated 
in the $x>w$ region.  We now write the Usadel equation for $f_1$ in this latter region. From Eq. (\ref{eq:linUsadelcompl}) we obtain:
\begin{equation}
    -\partial_{xx}^2f_1+\kappa_\omega^2 f_1=0\label{eq:lrtc}\; .
\end{equation}
This is the remarkable result: the triplet component $f_1$ is insensitive to the local exchange field and, once generated, can penetrate the right region over the long-range thermal length $\xi _{T}\approx
\sqrt{D_{F}/\pi T}$,  and does not oscillate in space , see Fig. \ref{fig3}b. If we choose $z$ as the spin quantization axis, then $f_3$ corresponds to triplet pairs with a zero total spin projection, $\left(\uparrow\downarrow+\downarrow\uparrow\right)$, whereas $f_1$ to pairs form by a combinations of $(\uparrow\uparrow)$ and $(\downarrow\downarrow)$ \cite{eschrig2011spin,hijano2022quasiparticle} (nematic case in terminology of
Ref.\cite{moor2015nematic}).

It is important to emphasize the   difference between triplet superconductivity in S/F
structures with an uniform and nonuniform magnetisation in F. In both cases,
the odd-frequency, triplet condensate component arises. However in the homogenoeus 
case the triplet component   penetrates into the ferromagnet only over short distances  from the interface where it coexists with the singlet, even-frequency,  component.   
In the inhomogeneous case certain triplet components of the condensate penetrate
into F over a large distance, see Fig. \ref{fig3}b,  so that in most part of ferromagnet only triplet superconductivity takes place. This is one of the first  firmly established
realization of the triplet superconductivity in solids. 
Indeed, the prediction of the long-range triplet component in SF hybrid
systems triggered intensive studies of the properties of this type of
superconductivity. A huge number of papers on the topic, both experimental and
theoretical , including several review articles \cite{buzdin2005proximity,bergeret2005odd,linder2015superconducting,eschrig2011spin,eschrig2015spin,linder2019odd}
have been published in the last two decades. 
Theoretical works focused on  the
long-range Josephson coupling \cite{volkov2003odd,volkov2005long,lofwander2005interplay,pajovic2006josephson,volkov2006odd,braude2007fully,houzet2007long,crouzy2007josephson,fominov2007josephson,sperstad2008josephson,halasz2009critical,konschelle2008nonsinusoidal,trifunovic2010long,volkov2010odd,alidoust2010spin,trifunovic2011josephson,trifunovic2011long,mel2012interference,richard2013superharmonic,hikino2013long,halterman2015charge,zyuzin2012magnon}, the peculiarities of the
density-of-states [86, 87] in S/F structures\cite{yokoyama2007manifestation,knevzevic2012signature},  and nonstationary effects in these structures \cite{ojajarvi2022dynamics,rabinovich2019resistive,bobkova2021dynamic}. Experimental detection  of the long-range triplet has been observed  in magnetic Josephson junctions and other S/F structures\cite{keizer2006spin,sosnin2006superconducting,khaire2010observation,anwar2010long,anwar2011inducing,anwar2012long,salikhov2009experimental,robinson2010controlled,visani2012equal,sanchez2022extremely,kalenkov2011triplet,klose2012optimization,blamire2014interface,caruso2019tuning,aguilar2020spin,satariano2021inverse,fermin2022superconducting}.

\section{Further developments and applications}

Long-range triplet superconductivity opened up novel opportunities of creating spin-polarized supercurrents, and hence a direct application of superconductors in spintronics\cite{eschrig2011spin,linder2015superconducting}.  
Currently, active research on triplet superconductivity in diffusive systems is continuing and new directions of this research are being opened. Above, we have quoted only a part of the publications on the subject under consideration. 
One of these  extensions of the  theory is the inclusion of spin-orbit coupling (SOC) in hybrid superconducting  diffusive structures. linear in momentum  SOC can be included in the form of a SU(2) vector potential, $\hat{\cal A}$\cite{bergeret2013singlet,bergeret2014spin}. In the strict diffusive limit the vector potential enters only as covariant derivatives  $\partial_k\rightarrow \tilde\partial_k=\partial_k-i[\hat A,.]$. This substitution in  Eq. (\ref{eq:linUsadelcompl} results in
\begin{equation}
    \hat \partial_{kk}^2\hat f=\partial_{kk}^2\hat f-2i[{\cal A}_k,\partial_k \hat f]-[{\cal A}_k, [{\cal A}_k,\hat f]]
\end{equation}
where summation over repeated indices is implied. Only triplet components enter the commutators in the second and third terms. These have exactly the same form as in the spin diffusion equation in the presence of SOC\cite{stanescu2007spin,duckheim2009dynamic}, establishing an interesting analogy between spin  and triplet Cooper pairs  diffusion. Indeed the second term describes  triplet precession, whereas the last term corresponds to relaxation due to SOC.  Both terms, in combination with a homogeneous exchange field, may lead to long-range triplet condensate \cite{bergeret2014spin}.  {It is also worth mentioning that the resolvent of the linearized Usadel equation corresponds to the Cooperon equation\cite{ilic2020unified}, which can be used to calculate the quantum corrections to conductivity, another topic for which Konstantin  B.  Efetov made important contributions\cite{efetov1983kinetics}.}

The analogy between spin and triplet Cooper pairs can be extended to describe magnetoelectric effects stemming from the charge-spin current. It was shown in Refs. \cite{bergeret2015theory,
konschelle2015theory} that known effects in the normal state, as the Edelstein and spin-galvanic effects \cite{aronov1991spin,edelstein1990spin,shen2014microscopic}
have their counterpart in superconducting systems due to the singlet-triplet conversion. 
In particular, Ref. \cite{konschelle2015theory} has established 
 a   direct connection between the latter effect and the appearance of an anomalous phase shift $\varphi_0$ in the current-phase relation of Josephson junctions with SOC, within the framework of the linearized Usadel equation\footnote{The exact Usadel equation in the presence of intrinsic SOC  has been obtained recently from the non-linear sigma model approach \cite{Virtanen_PhysRevB.105.224517}}.  $\varphi_0$-junctions have been realized experimentally\cite{szombati2016josephson,assouline2019spin,mayer2020gate,strambini2020josephson}. Other consequence of the singlet-triplet coupling is the so-called superconducting diode effect intensively studied nowdays\cite{ando2020observation,baumgartner2022effect,jeon2022zero,daido2022intrinsic,ilic2022theory,yuan2022supercurrent}.

Odd triplet superconductivity also plays a fundamental  role in the study of non-equilibrium properties of superconductor/ferromagnetic-insulator structures\cite{beckmann2016spin,
bergeret2018colloquium,HEIKKILA2019100540,hijano2022quasiparticle}. Such systems are being studied for applications in sensing\cite{heikkila2018thermoelectric}, cryogenic memories\cite{DeSimoni:2018}, and thermoelectricity\cite{machon2013nonlocal,ozaeta2014predicted}.

Summing up,  our group,  led by Konstantin Borisovich Efetov, predicted the odd-frequency triplet superconductivity in conventional S/F structures. Our findings 
opened up a new and very active research field. Kostya's role in our  research was not only limited to very constructive scientific discussions, but he also managed to create a rather friendly atmosphere in our
group, which stimulated fruitful cooperation and led to results that are
currently recognized as significant and new in the field of
superconductivity.

\section*{Acknowledgements}
\begin{acknowledgments}
FSB acknowledges financial support from Spanish AEI through project grant PID2020-114252GB-I00 (SPIRIT), the European Union’s Horizon 2020 Research and Innovation Framework Programme under Grant No. 800923 (SUPERTED), the A. v. Humboldt Foundation, and  the Basque Government through grant  IT-1591-22. 
\end{acknowledgments}

\bibliography{biblio}

\end{document}